# Thermophoretically driven capillary transport of nanofluid in a microchannel

Soumya Bandyopadhyay and Suman Chakraborty[*]

*Department of Mechanical Engineering, Indian Institute of Technology Kharagpur, West Bengal, India – 721302*

[*] Corresponding author. Tel: +91 3222282990.
E-mail address: suman@mech.iitkgp.ernet.in (S. Chakraborty).

HIGHLIGHTS

- Thermal gradient enhances nanoparticle deposition for a given nanofluid.
- The interplay between thermophoresis and diffusion governs particle migration.
- Thermal gradient and particle size enhance the rate of capillary transport.
- Optimum flow rates can be achieved for a given nanoparticle suspension.

## GRAPHICAL ABSTRACT

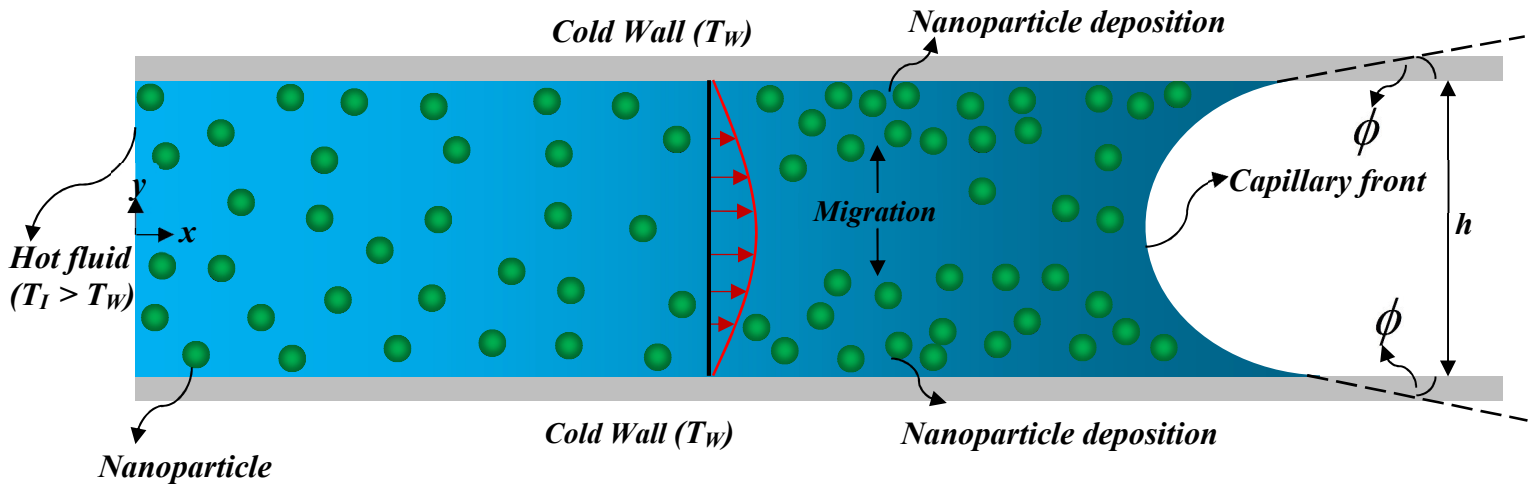

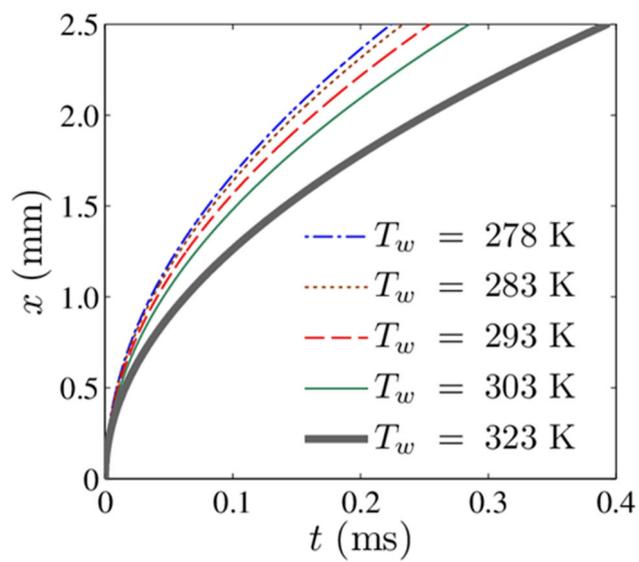

ABSTRACT

We investigate the interplay of thermophoretic force and interfacial tension on the capillary filling dynamics of a Newtonian nanofluid in a microchannel. In our model, we also consider an intricate thermofluidic coupling by taking the temperature dependence of viscosity aptly into account. This, in turn, determines the evolution of the viscous resistive force as the capillary front progresses, and presents an involved inter-connection between the driving thermophoretic force and the viscous resistive force. The two distinct regimes of particle transport in a fluid medium, delineated by particle size, are expounded to peruse the impact of imposed thermal gradients and particle size on particle retaining propensity of the nanofluid. Additionally, we witness a significant reduction in particle bearing proclivity of the nanofluid with enhancement in a thermal gradient. The results demonstrate the efficacy of the thermophoretic actuation towards the filling of narrow capillaries under the influence of a thermal gradient.

Keywords: Microchannel, Nanofluid, Suspension, Thermophoresis, Capillary transport

## 1. Introduction

Capillary-driven systems are ubiquitous in this physical world, with implications in diverse applications ranging from the transport of blood in the cardio-vascular pathways to the ascent of sap in plants occurring through the xylem and phloem tissues. Capillary transport has also been a widely pursued problem in the field of microfluidics [1-7]. Microfluidic transport has widespread importance, encompassing biomedical engineering [8-11] to flow-modulated cooling of electronic components [12,13] and also in chemical engineering [14,15]. Pressure-driven microfluidic transport has several disadvantages which include the paucity of precise experimental control, high pumping power requisites, and ample dispersion. This has led to the emergence of various other modes of transport which include electrosmosis [16-19], electrocapillarity [20], and electrokinetics [21]. Various state of the art models is available in this regard with an attempt to understand the physics of flow of Newtonian [22-25] as well as Non-Newtonian fluids [16,17]. Because of the immense diversity of the environments where capillary flows prevail, they are subjected to different physical forcing conditions arising from their surroundings. In this context, the detailed analysis of the physics of capillary flows attains paramount importance as it not only provides extremely useful theoretical insights to intricacies of the physical phenomenon by eliciting valuable scientific implications but also leads to the development of devices with widespread practical applications.

When there is a temperature gradient in a particle-laden fluid subjected to capillarity, the particles suspended in the fluid experience a force on account of thermophoresis, as expounded by Talbot et al. [26]. This thermophoretic force has been a driving factor behind the deposition of particles in a gaseous medium leading to commonly observed phenomena like blackening of a lantern. Based on particle sizes, different particles will experience different forces in presence of thermal gradients in a given flow field. As a consequence, thermophoresis results in varied migration rates of the particles, as delineated by Malvandi and Ganji [27], Guha and Samanta [28]. Hence, such thermal gradients which are prevalent in most of the modern day lab-on-a-chip devices may be pertinently utilized to regulate the resultant capillary dynamics. Additionally, temperature gradients existing in a system can be employed to separate different particles in a suspension. In the context of such systems, migration of nanoparticles eventually results in their preferential deposition determined by existing thermal fields. The separation of nanoparticles attains utmost importance not only in the context of achieving homogeneous suspensions but also in the design of particle retrieval systems essential for minimizing operational costs.

Choi and Eastman [29] coined the term "nanofluid" as a dilute mixture of particles, varying in size from 1 nm to 100 nm, suspended in a base fluid. Compared to base fluids, nanofluids inimitably possess superior thermophysical attributes like thermal conductivity [29]. Previously, Malvandi et al. [30] studied thermophoretic effects in nanofluids with a motivation to explore critical heat fluxes in boiling condition. Shiekholeslami et al. [31] delineated magnetohydrodynamic effects on natural convection with Cu-water nanofluid. Additional studies [32-34] have been conducted to expound intricate and interesting physics of nanofluids. Furthermore, researchers have investigated the enhancement of heat transfer in the convective flow of nanofluids in different geometries [35-37]. In microscale engineering systems like heat pipes and cooling components of electronic devices, where appreciable thermal gradients prevail, nanofluids can be utilized for their enhanced thermophysical properties. Consequently, the study of thermophoretic effects on the capillary transport of nanofluids in microfluidic confinements becomes immensely important.

This paper deals with the effects of thermophoresis on the capillary filling dynamics of nanofluids. The primary focus is to mathematically investigate the dynamics of capillary flow of nanofluids in the presence of thermal gradients, using a reduced order model, addressing the following: a) effect of particle size, b) influence of thermal gradient. Furthermore, the work is motivated by the requirement to achieve nanoparticle separations accompanied by a critical exegesis of the influence of pertinent parameters on nanoparticle deposition in microfluidic confinements. To the best of authors' knowledge, no comprehensive study thus far has been conducted to explore capillary dynamics in presence of thermophoretic effects. The present theoretical findings are expected to impel the researchers in performing experiments with nanofluids to explicate thermophoretic effects in microscale confinements. In the following section, the mathematical model used to illustrate the above-mentioned physical aspects is discussed in detail, along with the relevant fundamental parameters that implicitly influence the interaction of the various important facets controlling the physics of flow.

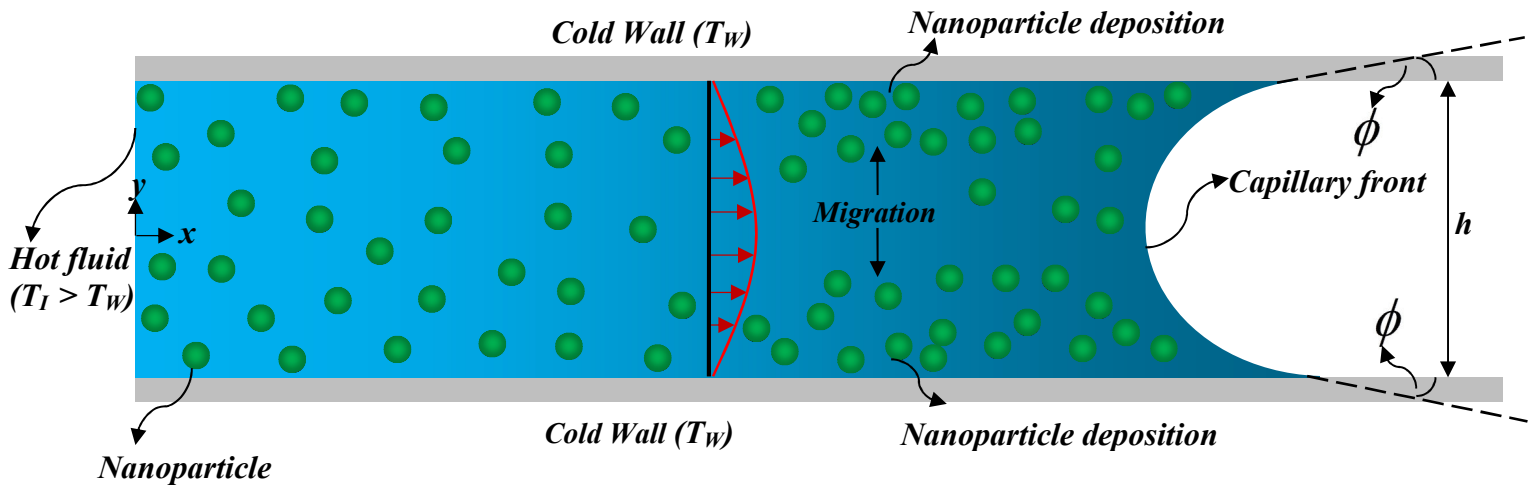

**Fig. 1.** Physical model depicting the layout of the two-dimensional microchannel and the coordinate system

## 2. Mathematical formulation and numerical procedure

*2.1. Physical problem*

We consider parallel plate microchannel geometry, with the two plates being separated by a distance $h$ which is equal to 500 μm. The channel is assumed to have a width $b$ perpendicular to the plane of the diagram, such that $b>>h$. The fluid has initially traversed a perfectly insulated section which is long enough for the flow to become hydro-dynamically fully developed. We set our origin at the center of the channel in the beginning of the non-insulated section (see Fig. 1), with $y$ axis running along the transverse direction and $x$ axis along the length of the channel. The static contact angle at the solid-liquid-gas (air) interface is denoted by $\phi$. There will be a transition of the fully developed velocity profile to a meniscus traction regime near the interface via a transition region. However, the length of the meniscus traction regime is very small compared to the length of the fully developed region and hence, is neglected in this model.

At the inlet of the non-insulated section, we have a uniform temperature ($T_l$), concentration ($C_o$) and fully developed velocity field. The walls are cooled to a constant temperature ($T_w$). The nanofluid under consideration conceived as a suspension of nanoparticles comprises two discrete phases-1) The liquid phase (water) 2) The particle phase ($SiO_2$). One of the most important considerations of the present model is that there is uni-directional coupling (i.e. the particle motion is determined by the flow field not the other way around) and there will be no mutual interaction between the particles. A similar model has been adopted by Guha and Samanta [28], Chein and Liao [38]. The potency of this assumption is attributed to the fact that we will be dealing with nanofluids of extremely dilute concentrations ($C \sim 10^{-3}$ M) and nanoparticles of sizes ($d \sim 1$ nm), resulting in low volume fraction ($\varphi \sim 0.001\%$). In the current study, gravity, inter-particle forces, magnetic and electrostatic forces on the particle are all neglected. The temperature dependence of viscosity is suitably taken into account in the present model through its dependence on the mean temperature at a given cross-section. Consequently, it is a function of the axial coordinate (*x*) only. Except for viscosity, other thermo-physical properties of the fluids are evaluated at the average of the maximum and minimum temperatures. The temperature dependence of thermophysical properties of the fluid is given by Eq. (A.1-4). The thermophysical property of a given nanofluid like thermal conductivity, heat capacity, dynamic viscosity was computed as a function of the particle volume fraction ($\varphi$) by He et al. [39]. Later, Ganguly et al. [31] incorporated established models to obtain pertinent thermophysical properties of the nanofluids in their study. However, in the limit of low volume fractions, the thermophysical property of a nanofluid becomes equivalent to the corresponding property of the fluid phase. Hence, in the present analysis, the nanofluid properties have been replaced by the analogous properties of the fluid phase.

Here, we apply a reduced order model [40] that is commonly used for evaluating the capillary filling distance as a function of time. This model has its inception from the early works of Lucas [41] and Washburn [42]. By dynamics of the capillary transport following this model, what we intend to study is the average position (*x*) of the capillary front at a time *t*. The equation of motion for the capillary advancement taking the appropriate direction of the forces, following Newton's second law of motion, may be expressed as (neglecting inertial forces consistent with a microfluidic paradigm):

$$F_{ST} + F_V(x) + F_T(x) = 0 \tag{1}$$

Where $F_{ST}$, $F_V(x)$ and $F_T(x)$ denote forces due to surface tension, viscous resistance at the walls and thermophoresis respectively.

The force due to surface tension is modeled under the consideration of static contact angle and particle concentration dependence of surface tension coefficient on interfacial particle concentration. We have also tested for results with dynamic contact angle and no qualitative changes in the results have been observed. That is why, for simplicity yet without sacrificing the essential physics that we intend to capture, we account for static contact angle only. $F_{ST}$, thus, can be expressed as:

$$F_{ST} = \sigma P \cos\phi \tag{2}$$

where σ stands for the surface tension coefficient of the liquid, *P* denotes the perimeter of the cross-section of the microchannel and ϕ stands for the static contact angle of the liquid-channel interface and is chosen as $40°$ in the present study. The computation of $F_V(x)$ necessitates the requirement of knowledge of the velocity gradients in the flow. For this purpose, the following expression of the axial velocity variations, considering hydrodynamically fully developed flow in a local sense, is used:

$$\frac{u(x,y)}{u_{avg}(x)} = \frac{3}{2}(1 - 4\overline{y}^2) \tag{3}$$

$$\frac{u_{avg}(0)}{u_{avg}(x)} = \frac{\mu_l(x)}{\mu_l(0)} \tag{4}$$

$$F_V(x) = \frac{12\mu_l(x)\dot{x}xb}{h} \tag{5}$$

Here, $u_{avg}(x)$ refers to the average inlet velocity at any particular cross-section, $\dot{x}$ stands for the velocity of the capillary front, where *x* is the instantaneous position of the capillary front from the origin of the reference coordinate system and $\mu_l(x)$ is the dynamic viscosity of the fluid. The thermophoretic force acting on a particle due to thermal gradients and concentration gradients, as formulated by Talbot et al. [26], is integrated over the volume to obtain the net thermophoretic force:

$$F_T(x) = -\int_0^x \int_{-\frac{h}{2}}^{\frac{h}{2}} D_{T,P}(x) \frac{\partial T}{\partial x} \frac{C(x,y)}{T(x,y)} b\,dx\,dy \tag{6}$$

$$D_{T,P}(x) = \frac{6\pi d \mu_l(x)^2 C_s (K + C_t Kn)}{\rho_l (1 + 3C_m Kn)(1 + 2K + 2C_t Kn)} \tag{7}$$

$C_m$, $C_s$, $C_t$ are the momentum exchange coefficient, thermal slip coefficient, and the temperature jump coefficient, and are taken to be equal to 1.146, 1.147 and 2.18, respectively [39]. $K$ is the ratio of the base fluid thermal conductivity $k_l$, computed from Eq. (A.3), and the particle thermal conductivity $k_p$. In the present working temperature range of 278 K to 323 K, $k_p$ is treated as a constant which equals 1.38 W/mK. $Kn$ denotes the Knudsen number and is defined as $Kn = 2\lambda/d$, where $\lambda$ stands for the mean free path of the working fluid and $d$ denotes the diameter of the nanoparticles.

Estimation of $F_T(x)$ necessitates the knowledge of temperature distribution $T(x,y)$ and resulting particle distribution in the fluid $C(x,y)$. For the calculation of temperature field, quasi-steady-state approximation and negligible axial conduction in conjunction with no viscous dissipation and work-transfer are considered. The effect of variation of dynamic viscosity with temperature is also accounted for. In the entire study, $T_l$ is taken to be 323K and $T_w$ is varied from 278 K to 323 K. The energy equation now can be written in terms of the non-dimensional temperature $\theta(x,y)$ as:

$$\theta(x,y) = \frac{T(x,y) - T_w}{T_w} \tag{8a}$$

$$\overline{T}(x) = \int_{-\frac{h}{2}}^{+\frac{h}{2}} T(x,y)\,dy \tag{8b}$$

$$\frac{u(x,y)}{u_{avg}(0)} \frac{\partial \theta}{\partial \overline{x}} = \frac{1}{Pe} \frac{\partial^2 \theta}{\partial \overline{y}^2} \tag{9}$$

$Pe$ stands for the Peclet number of the flow based on inlet Reynolds number $Re$ and Prandtl number $Pr$. $\overline{x}$, $\overline{y}$ and $\overline{l}$ are the non-dimensionalized versions of $x$, $y$, and $l$ after being scaled by h. The boundary conditions for Eq. (9) are prescribed as follows:

$$\theta\big|_{\overline{y}=-1/2} = 0 \quad \forall \quad 0 < \overline{x} < \overline{l} \tag{10a}$$

$$\theta\big|_{\overline{y}=+1/2} = 0 \quad \forall \quad 0 < \overline{x} < \overline{l} \tag{10b}$$

$$\theta\big|_{\bar{x}=0} = \theta_I \qquad \forall \qquad -\frac{1}{2} < \bar{y} < \frac{1}{2} \qquad (10c)$$

For the estimation of the particle concentration in the fluid, we have adopted a quasi-steady-state approximation along with minimal axial diffusion. Particle velocity in the convective term is replaced by the axial velocity of the fluid, as the particle size is in the order of nm. The thermophoretic effect on particle migration has been accounted by the introduction of an additional term in the species conservation equation [28,38]. The concentration of particles near the wall in the fluid is taken as zero, as we have assumed that particles adhere to the wall as they impinge on the wall. This assumption is ideally applicable for particles of diameter less than 2 nm but has been used for particles as large as 10 nm by Chein and Liao [38]. In the present study, we have confined ourselves to the maximum particle size of 5 nm. The species conservation equation can be written in terms of non-dimensional concentration $\bar{C}(x,y)$ as:

$$\bar{C}(x,y) = \frac{C(x,y)}{C_0} \qquad (11)$$

$$\frac{u(x,y)}{u_{avg}(0)} \frac{\partial \bar{C}}{\partial \bar{x}} = \frac{1}{Pe_c} \frac{\partial^2 \bar{C}}{\partial \bar{y}^2} + \frac{K_{T,P}}{Re} \frac{\partial}{\partial \bar{y}}\left( \frac{\bar{C}}{1+\theta} \frac{\partial \theta}{\partial \bar{y}} \right) \qquad (12)$$

Here $C_0$ is the concentration of particles at the inlet of the channel and is taken to be 1 mM/m$^3$ for the study. The particle Peclet number $Pe_c$ is based on $K_{T,P}$ in this equation stands for the thermophoretic coefficient for the concentration term which is denoted by:

$$K_{T,P} = \frac{2C_s C_c (K + C_t Kn)}{(1+3C_m Kn)(1+2K+2C_t Kn)} \qquad (13)$$

Where $C_c$ is the equivalent Cunningham correction factor in the present case, similar to the formulation used by Chein and Liao [38].

$$C_c = 1 + Kn \left[ 1.257 + 0.4 \, exp\left(\frac{-1.1}{Kn}\right) \right] \qquad (14)$$

The particle diffusivity $D$, evoking Stokes Einstein theory, may be estimated as:

$$D = \frac{C_c}{3\pi d \mu_{avg}} k_B T_{avg} \qquad (15)$$

The Schmidt number takes the form:

$$Sc = \frac{\mu_{avg}}{\rho_l D} \qquad (16)$$

In Eq. (15), $k_B$ stands for the Boltzmann constant and is equal to $1.38 \times 10^{-23}$. $\mu_{avg}$ and $\rho_l$ are the dynamic viscosity and density at $T_{avg}$ which is the average of the maximum and minimum temperatures. The relevant boundary conditions are as follows:

$$\overline{C}\big|_{\overline{y}=-1/2} = 0 \quad \forall \quad 0 < \overline{x} < \overline{l} \tag{17a}$$

$$\overline{C}\big|_{\overline{y}=+1/2} = 0 \quad \forall \quad 0 < \overline{x} < \overline{l} \tag{17b}$$

$$\overline{C}\big|_{\overline{x}=0} = 1 \quad \forall \quad -\frac{1}{2} < \overline{y} < \frac{1}{2} \tag{17c}$$

Having computed all the forces, Eq. (1) can be rewritten as:

$$\sigma P \cos\phi + F_T(x) - \frac{12\mu_l(x)\dot{x}xb}{h} = 0 \tag{18}$$

As $b \to \infty$, Eq. (18) takes the form:

$$\int_0^x \frac{dx}{p(x)} = t \tag{19}$$

where,

$$p(x) = \frac{\sigma h \cos\phi}{6x\mu_l(x)} + \frac{C_0 F(x) h^2}{12x\mu_l(x)} \tag{20}$$

Here $F(x)$ is the scaled thermophoretic force and is expressed as:

$$F(x) = \frac{F_T(x)}{C_0 bh} \tag{21}$$

Where $F_T(x)$ is given by Eq. (6).

## 3. Numerical Method

Eq. (9) and Eq. (11) with the boundary conditions given by Eq. (10a-c) and Eq. (17a-c) represent a set of coupled non-linear partial differential equations which has been solved using a forward-marched finite difference scheme. First, Eq. (9) is solved numerically along with relevant boundary conditions as given by Eq. (10a-c). The mean temperature $\overline{T}(x)$ subsequently computed by Eq. (8b) yields the viscosity from Eq. (A.1). As a consequence, the solution of the energy equation yields the temperature variation as well as the axial variation of viscosity. Next, Eq. (11) is solved numerically with pertinent boundary conditions as given by Eq. (17a-c), evoking the computed temperature gradients and the viscosity variation, from Eq. (9). The computed temperature and concentration fields are subsequently utilized in Eq. (6) to calculate the net thermophoretic force. Finally, Eq. (19) is numerically integrated to yield the capillary dynamics. A Matlab code has been used to numerically solve the system of coupled partial differential equations followed by computation of thermophoretic force and subsequently obtaining the capillary dynamics. The accuracy and validation of the code have been discussed in the following section.

## 4. Grid testing and validation of current solution scheme

The grid dependency test is conducted by computing the transverse concentration profiles at $\bar{x} = 1.25$ corresponding to d = 3 nm and $T_w$ = 278 K (See Fig. 2a). The average non-dimensional concentration ($\bar{C}_{avg}$), corresponding to a particular grid size is calculated and the results are shown in Table 1. The grid size is varied from 20 × 5 000 till 70 × 80 000. Based on the tabulated results, $\bar{C}_{avg}$ does not vary appreciably beyond an optimum grid size of 40 × 35 000 and hence it has been selected for computational purposes. The validation of the numerical technique has been done by treating viscosity as a constant in the current model and by using the same parameters as used in [38] for air-SiO$_2$ nano-particle suspension. Fig. 2b depicts the accuracy of the current solution scheme relative to the existing numerical results.

**Table 1**

Comparison of the average non-dimensional concentration $\bar{C}_{avg}$ for different grids at $\bar{x} = 1.25$ with d = 3 nm at $T_w$ = 278 K

$$\bar{C}_{avg} = \int_{-\frac{1}{2}}^{+\frac{1}{2}} \bar{C}(\bar{x} = 1.25, \bar{y}) d\bar{y}$$

| Mesh size | 20 × 5000 | 30 × 20 000 | 40 × 35 000 | 50 × 80 000 | 60 × 80 000 | 70 × 80 000 |
|---|---|---|---|---|---|---|
| $\bar{C}_{avg}$ | 0.6735 | 0.6811 | 0.6850 | 0.6873 | 0.6888 | 0.6898 |

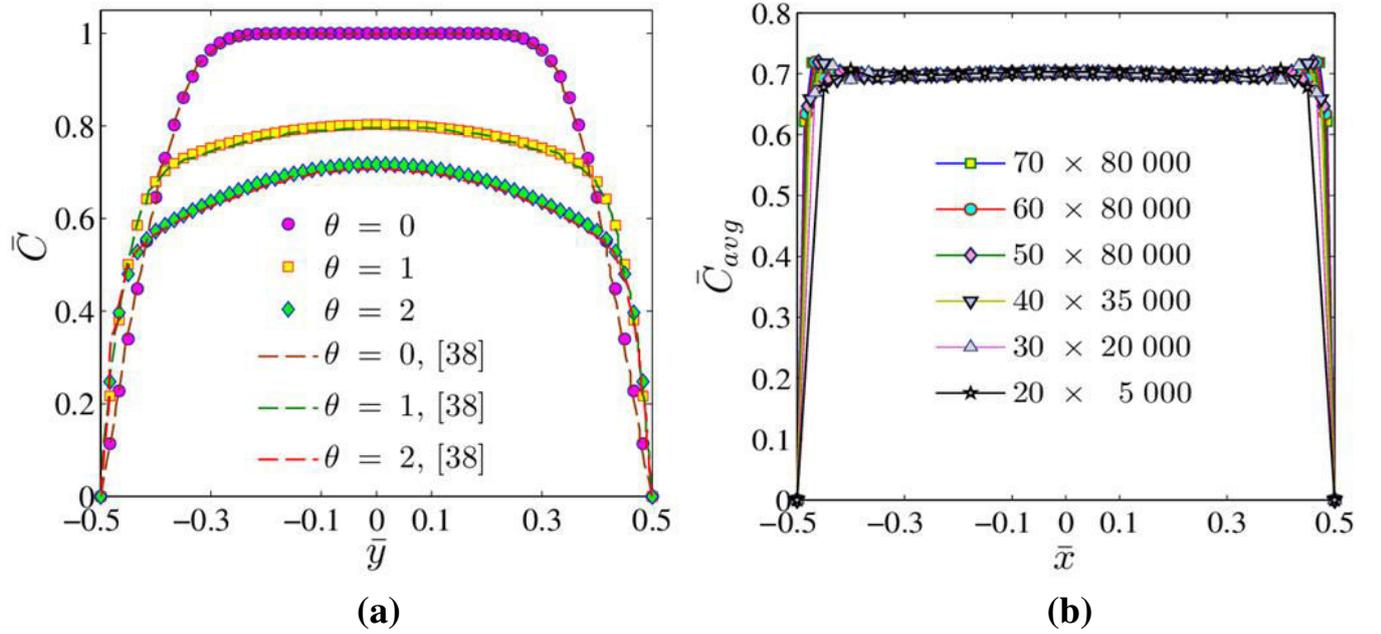

**Fig. 2.** (a) Comparison of the transverse concentration profile between present results and numerical results reported by Chein and Liao [38] air-SiO$_2$ nanoparticle suspension at different inlet temperatures; (b) Grid dependency test with different grid sizes.

## 5. Results and discussion

Fig. 3a and Fig. 3b depicts the axial variation of the mean temperature and viscosity of the fluid, respectively. As the wall temperature approaches the temperature of the incoming fluid, there is a decrement in the length of the thermal developing region which affects the axial variation of the dynamic viscosity. Consequently, the mean temperature of the fluid decreases and the fluid attains the wall temperature. The viscosity of the fluid increases in the thermally developing region and becomes constant and viscosity variations gradually lose prominence as the temperature of the incoming fluid approached wall temperature. In the present scenario, thermal gradients are appreciable up to $\bar{x}=2$ for the maximum temperature gradient case. Temperature and subsequently the particle concentration profiles are of significance up to $\bar{x}=5$, as beyond that thermophoretic effects lose prominence.

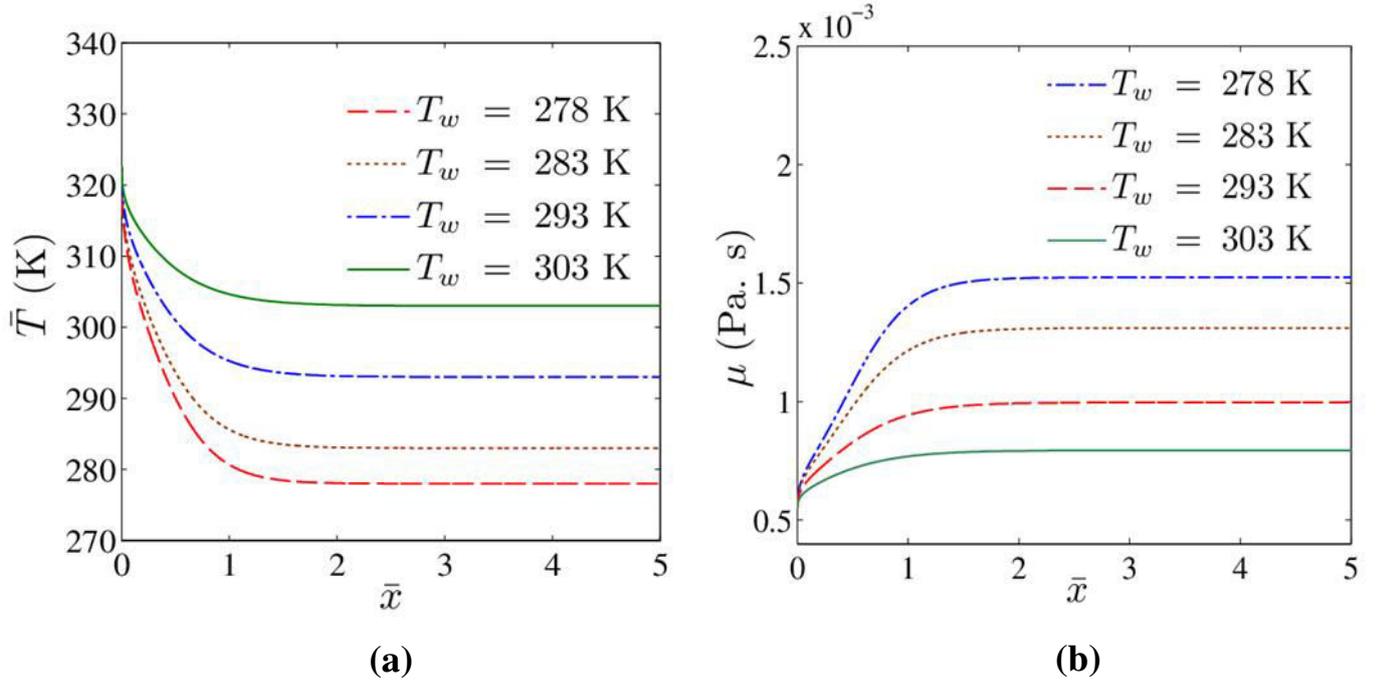

**Fig. 3.** (a) Axial variation of the mean temperature of the fluid, for different wall temperatures; (b) Axial variation of the viscosity of the fluid for different wall temperatures.

Due to the inherent symmetry of the problem in terms of the boundary conditions, the particle concentration profiles are symmetric about the central axis of the microchannel; see Fig. 4a. The concentration gradient near the wall decreases as we traverse along the axis of the microchannel due to the gradual development of the particle concentration profile. Increased contact length of the fluid with the wall effects in enhanced thermophoretic particle deposition on the channel walls, inhibiting particle concentrations at a given transverse location as we traverse axially. Concentration profiles for a given nanofluid in a specified confinement, subjected to a particular temperature field, clearly depict that with a gradual increase in length along the central axis, accretion in thermophoretic effects diminishes. This phenomenon is ascribed to the eventual depletion of thermal gradients, as the capillary front progresses.

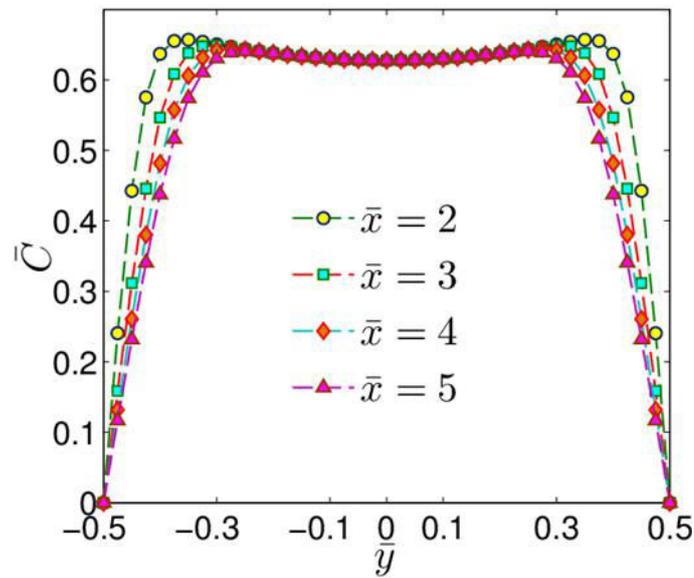

(a)

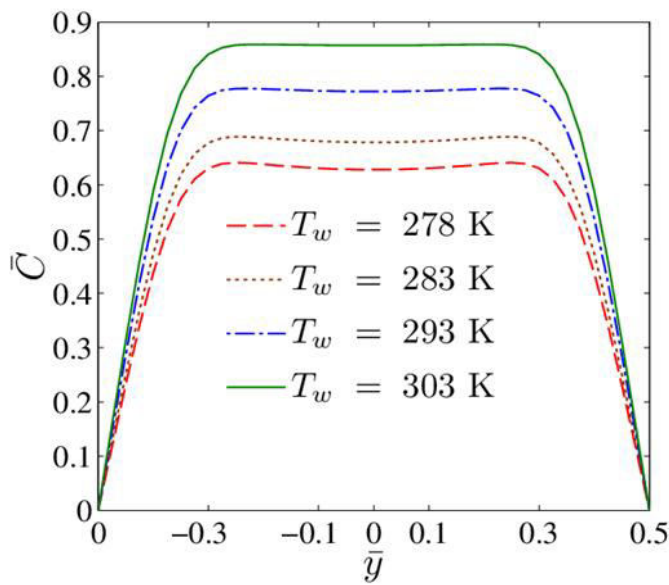

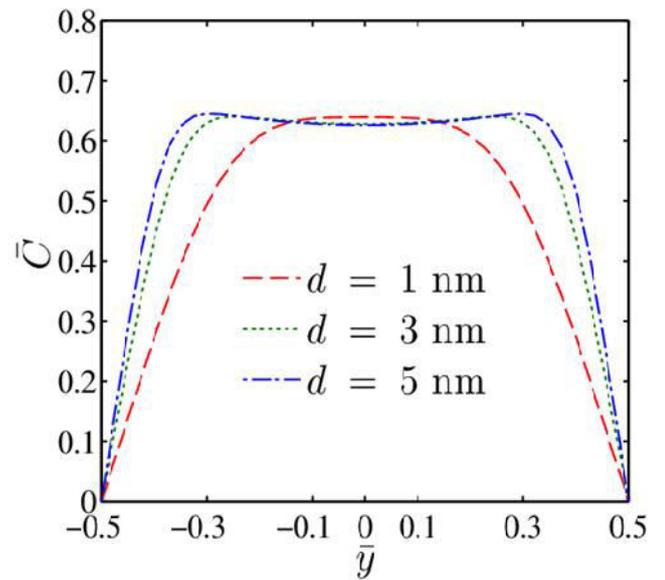

(b) (c)

**Fig. 4.** (a) Concentration profiles at various axial locations, at $T_w$ = 278 K, with $d$ = 3 nm; (b) Concentration profiles at $\bar{x} = 5$ for $d$ = 3 nm, at different wall temperatures; (c) Concentration profiles at $\bar{x} = 5$ for varied particle size, at $T_w$ = 278 K

Fig. 4b depicts the variations in the concentration profiles for different wall temperatures, corresponding to $\bar{x} = 5$, for $d$ = 3 nm. Thermophoretic effects become subdued with the increase in the wall temperature due to the reduction of axial temperature gradients in the system and particle diffusion becomes the dominant phenomenon. As a consequence, thermophoretic particle deposition reduces, leading to higher accumulation of particles near the central axis of the microchannel. The reliance of the particle-retaining capacity of the nanofluid on the interplay between thermophoretic and Brownian motion forces implies that, in a given microfluidic confinement, particle deposition can be pertinently tailored by modulating the externally imposed thermal gradients. Consequently, this attains utmost importance in nanofluid based microscale engineering applications where particle deposition exceeding a given threshold will be detrimental to the performance of the system.

For a specified wall temperature ($T_w$ = 278 K), concentration profiles at $\bar{x} = 5$ are depicted for different particle sizes ranging from 1 nm to 5 nm, in Fig. 4c. The particle size is reflected in the Schmidt number in the species conservation equation. Fig. 4c shows that increase in Schmidt number ($Sc$) undermines the diffusion phenomena compared to thermophoresis and thereby leads to the increase in concentration gradients near the wall for larger particles. In the limiting case when thermophoresis is negligible for lower particle sizes, diffusion is the predominant phenomenon. It can be inferred that smaller nanoparticle deposition is driven by Brownian forces, contrary to larger nanoparticle deposition, which is dictated by thermophoresis. The transition from molecular diffusion dominated particle deposition to the thermophoretically dominated regime, in turn, determines the particle-bearing capacity of the suspension. Consequently, the particle-bearing efficacy of a given nanofluid can be suitably exploited to design ancillary particle retrieval systems in diverse engineering applications deploying nanofluids in microscale confinements.

The variation of scaled thermophoretic force as a function of the axial coordinate is analyzed for particles of different sizes at a fixed wall temperature (see Fig. 5a). The thermophoretic force exhibits an increasing trend up to $\bar{x} = 2$ and then attains saturation. This can be explained by the fact that the flow becomes approximately thermally fully developed after $\bar{x} = 2$. Thermophoretic force increases with increase in particle size for a given thermal gradient; this is in concurrence with the results depicted before in this paper. The variation of normalized thermophoretic force as a function of the axial coordinate is analyzed for different wall temperatures for a fixed particle size of 3 nm (see Fig. 5b). The magnitude of the thermophoretic force falls with the rise in the wall temperature. This is solely attributed to the reduction in the temperature gradients in the flow. Moreover, with an increase in the wall temperature, there is a reduction of the length, over which there is an appreciable variation of thermophoretic force. This is corroborated by the computed thermal field, discussed earlier and depicted in Fig. 3a.

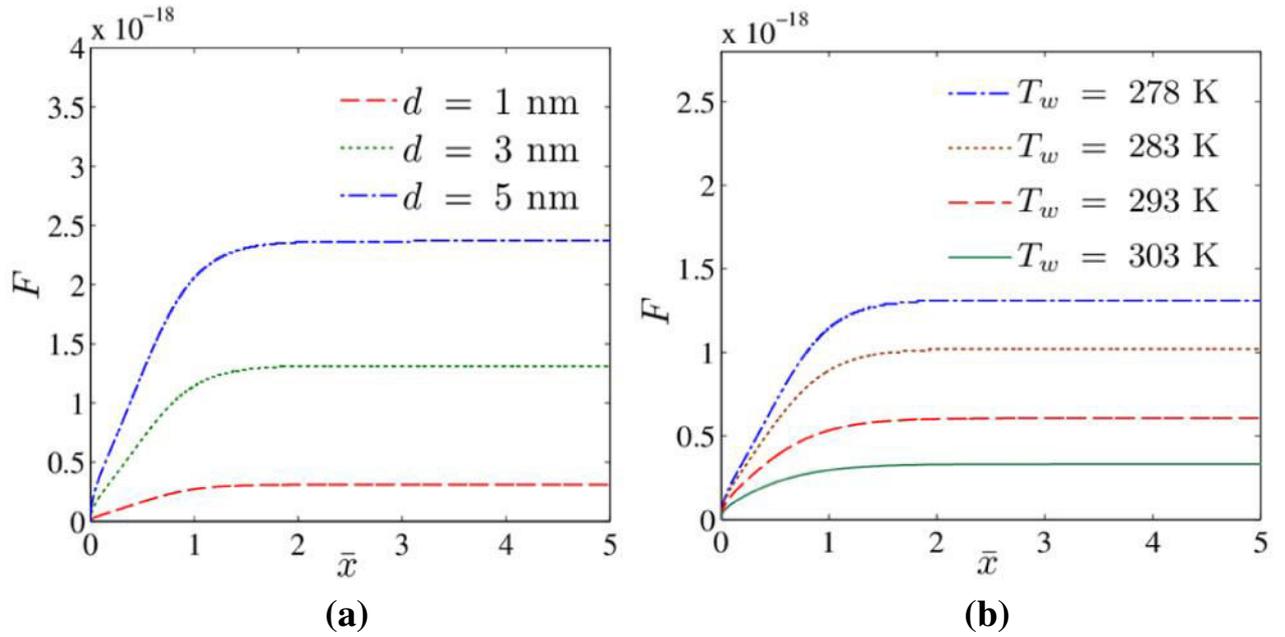

**Fig. 5.** (a) Variation of scaled thermophoretic force with axial position for particles of different sizes at $T_w$ = 278 K; (b) Variation of scaled thermophoretic force with axial position for $d$ = 3 nm at different wall temperatures.

Fig. 6a depicts the capillary filling characteristics, as a function of the particle size. With the increase in particle size, the thermophoretic force increases and consequently, enhancing the flow rate of nanoparticle suspension, as observed till $\bar{x} = 5$. It can be noted that the increase in capillary filling rate is significantly higher for lower particle size. This is primarily attributed to the relatively high increase in the thermophoretic force at lower particle size. In this context, the influence of particle size on capillary dynamics can be utilized to achieve separation of nanoparticles in a heterogeneous nanofluid. Fig. 6b shows the variation in capillarity, as a function of wall temperature. As the wall temperature approaches the incoming fluid temperature, flowrates reduce. This is in consensus with the gradual reduction of thermophoretic force with increase in wall temperature for a given nanofluid, as discussed earlier. Thermophoresis clearly enhances the axial rate of transport as compared to purely surface tension driven transport ($T_w = T_I$); (see Fig. 6b). It is worthwhile to note here, that beyond a particular wall temperature for identical inlet conditions, the rate of capillary transport does not increase significantly. Hence, optimum flow rates do exist for a particular incoming nanofluid in a given microfluidic confinement.

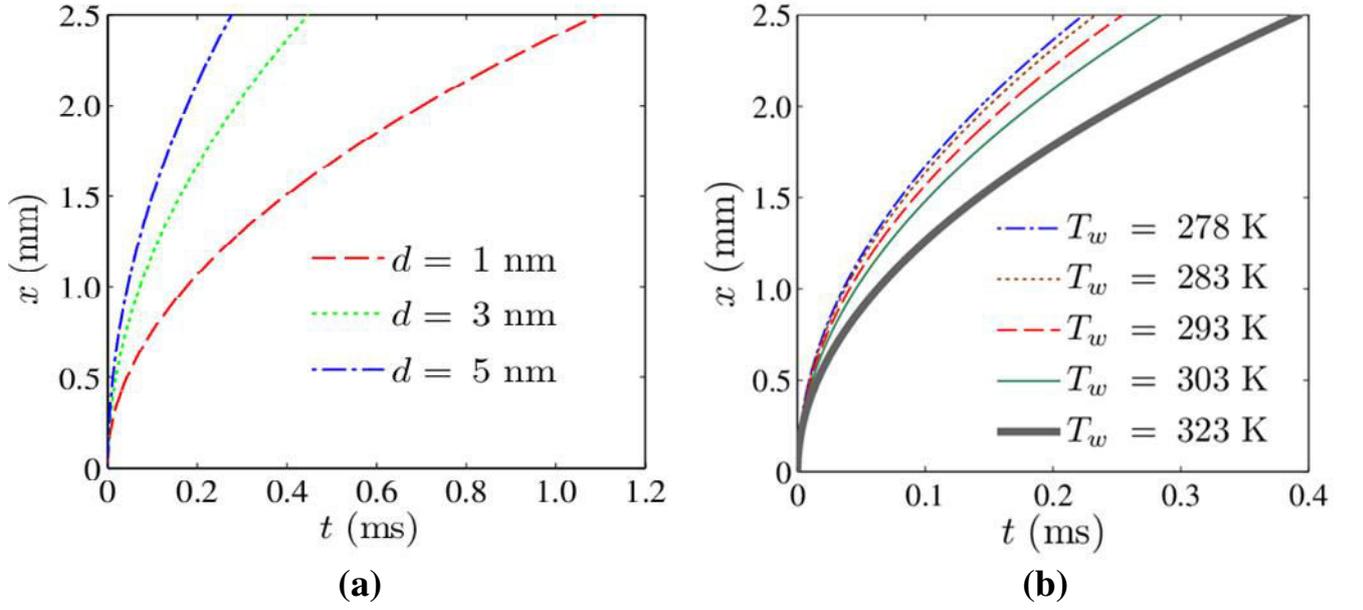

**Fig. 6.** (a) Variation of scaled thermophoretic force with axial position for particles of different sizes at $T_w$ = 278 K; (b) Variation of scaled thermophoretic force with axial position for $d$ = 3 nm at different wall temperatures. The solid thick line ($T_I = T_w$ = 323 K) in the figure corresponds to purely surface tension driven capillary transport.

## 6. Conclusions

A general transport model of the nanofluid has been developed in presence of thermophoretic force for explicating particle migration in a fluid medium and eventually delineating the evolving capillary dynamics. A uni-directional coupling between the energy equation and species conservation equation has been aptly considered, with the incorporation of the effect of thermophoresis in the species conservation equation. The most important feature of the present mathematical model is the judicious inclusion of the temperature dependence of viscosity. Findings reveal the existence of two discrete phenomena governing particle transport, dictated by the particle size for a prescribed thermal gradient, namely- diffusion and thermophoresis; the interplay between them influences particle deposition, thereby modulating the particle retaining potential of the nanofluid. In conclusion, thermal gradients in a confinement exacerbate the particle-bearing capability of a particular nanofluid. Moreover, the current model elucidates the effects of thermal gradients and particle sizes on the transport of the nanofluid. One of the most important outcomes of this mathematical model is that in practice, nanoparticle separation can be achieved by employing thermal gradients in a heterogeneous suspension. It is explicitly demonstrated that thermophoresis clearly enhances the rate of transport relative to purely surface tension driven capillary transport. Furthermore, for a given particle size, thermophoretic force, and consequently, the resultant capillary dynamics get influenced by the presence of thermal gradients in the system. In this respect, by the appropriate control of channel wall temperature, one can attain optimum rates of transport, reducing the energy costs.

# Appendix A

The thermophysical properties of water obtained from the ASHRAE Handbook [43] were used to fit a curve for computing the properties at any given temperature $T$ in our working range (278 K – 323 K).

*-Thermophysical properties of water*

$$\mu_l = -0.0000000064998705T^3 + 0.00000626195093807T^2 - 0.00202099183202047T + 0.2190624024917910 \tag{1}$$

$$\rho_l = 0.0000314569536447T^3 - 0.0330124263165957T^2 + 11.0386601961225007T - 193.285493540877000 \tag{2}$$

$$k_l = 0.00000022710464T^4 - 0.00027624146191T^3 + 0.125951240647987T^2 - 25.51100962753100T + 1,937.27799984267000 \tag{3}$$

$$Pr = -0.00007726338280T^3 + 0.07311004221889T^2 - 23.14366046554400T + 2,455.08343412100000 \tag{4}$$

# Conflict of interest

The authors of this manuscript have no conflict of interest.